\documentclass[aps,prd,reprint,superscriptaddress,nofootinbib,preprintnumbers,longbibliography,showkeys]{revtex4-2}

\usepackage{ORCIDinREVTeX}

\usepackage[T1]{fontenc}
\usepackage{slashed}
\usepackage{amsmath,amssymb,dsfont,epsfig,graphicx}
\usepackage{ragged2e}
\usepackage{float}
\usepackage[utf8]{inputenc}
\usepackage{enumitem}
\usepackage{xcolor}
\usepackage[colorlinks=true,urlcolor=blue
,anchorcolor=blue
,citecolor=purple
,filecolor=blue
,linkcolor=blue
,menucolor=blue
,pagecolor=blue
,linktocpage=true
,pdfproducer=medialab
,pdfa=true
]{hyperref}
\allowdisplaybreaks
\usepackage{mathtools}

\numberwithin{equation}{section}

\newcommand{\nc}{\newcommand}
\nc{\non}{\nonumber}
\nc{\hc}{\hbox {h.c.}}
\nc{\noi}{\noindent}
\nc{\barx}{\bar{x}}
\nc{\pbarn}{\;\hbox {pb}}
\nc{\fbarn}{\;\hbox {fb}} 

\nc{\hsp}{\hspace{0.5cm}}
\nc{\lsp}{\hspace{1cm}}
\nc{\Lsp}{\hspace{2cm}}
\nc{\LLsp}{\lsp\lsp}
\nc{\lra}{\longrightarrow}
\nc{\p}{\prime}
\nc{\sgn}{\text{sgn}}
\nc{\ph}{\varphi}
\nc{\op}{{\cal O}}

\def\eq#1{Eq.~(\ref{#1})}
\def\fig#1{Fig.~\ref{#1}}
\def\sec#1{Sec.~\ref{#1}}

\nc{\beq}{\begin{equation}}  \nc{\eeq}{\end{equation}}
\nc{\bea}{\begin{eqnarray}}  \nc{\eea}{\end{eqnarray}}
\nc{\baa}{\begin{array}}     \nc{\eaa}{\end{array}}
\nc{\bit}{\begin{itemize}}   \nc{\eit}{\end{itemize}}
\nc{\ben}{\begin{enumerate}} \nc{\een}{\end{enumerate}}
\nc{\bce}{\begin{center}}    \nc{\ece}{\end{center}}
\nc{\bpm}{\begin{pmatrix}}   \nc{\epm}{\end{pmatrix}}
\nc{\bvt}{\begin{verbatim}}  \nc{\evt}{\end{verbatim}}

\def\lsim{\mathrel{\raise.3ex\hbox{$<$\kern-.75em\lower1ex\hbox{$\sim$}}}}
\def\gsim{\mathrel{\raise.3ex\hbox{$>$\kern-.75em\lower1ex\hbox{$\sim$}}}}

\def\udots{\mathinner{\mkern1mu\raise1pt\vbox{\kern7pt\hbox{.}}\mkern2mu\raise4pt\hbox{.}\mkern2mu\raise7pt\hbox{.}\mkern1mu}}

\def\mev{\;\hbox{MeV}}
\def\gev{\;\hbox{GeV}}
\def\tev{\;\hbox{TeV}}

\def\mpl{M_{\text{Pl}}}

\definecolor{agray}{rgb}{0.95, 0.95, 0.99}

\nc{\ctw}{\cos\theta_{\textsc w}}
\nc{\stw}{\sin\theta_{\textsc w}}
\nc{\cwsq}{\cos^2\theta_{\textsc w}}
\nc{\swsq}{\sin^2\theta_{\textsc w}}
\definecolor{cred}{rgb}{0.6, 0.0, 0.0}
\setlist[itemize]{itemsep=0em, topsep=0.3em}

\begin{document}
\title{Ultraviolet freeze-in dark matter through the dilaton portal}
\author{Aqeel Ahmed}
\orcid{0000-0002-2907-2433}
\email{aqeel.ahmed@mpi-hd.mpg.de}
\affiliation{Max-Planck-Institut für Kernphysik (MPIK), Saupfercheckweg 1, 69117 Heidelberg, Germany}

\author{Saereh Najjari}
\orcid{0000-0003-3053-3759}
\email{snajjari@uni-mainz.de}
\affiliation{PRISMA+ Cluster of Excellence \& Mainz Institute for Theoretical Physics,\\ Johannes Gutenberg University, 55099 Mainz, Germany}

\date{\today}

\begin{abstract}
We study a class of models in which the Standard Model (SM) and dark matter (DM) belong to a conformal/scale-invariant theory at high energies. Scale invariance is spontaneously broken at scale $f$, giving rise to a dilaton as the corresponding Goldstone boson. In the low energy theory, we assume that DM interacts with the SM solely through the dilaton portal, which is suppressed by the conformal breaking scale $f$. For $f\gg{\rm TeV}$, the portal interactions are extremely weak, resulting in DM not being in thermal equilibrium with the SM. Thus, ultraviolet freeze-in production of DM occurs through the dilaton portal, being most effective at the maximum temperature of the SM bath. The temperature evolution is greatly impacted by the reheating dynamics, which we parametrize using a general equation of state $w$ and temperature at the end of reheating $T_{\rm rh}$. We analyze the implications of the reheating dynamics for DM production in this framework and identify regions of parameter space that result in the observed DM relic abundance for a wide range of DM masses and reheating temperatures for scalar, vector, or fermion DM.
\end{abstract}

\keywords{Dark Matter, Dilaton Portal,  Freeze-in DM, Scale Invariance, Beyond the Standard Model, Reheating, Nonstandard cosmology}
\preprint{MITP-21-043}

\maketitle
\tableofcontents

\section{Introduction}
\label{sec:intro}
The Standard Model (SM) of particle physics presents a remarkably successful description of visible matter constituents and their interactions. 
However, there remain some puzzles that are unanswered within the SM framework including the nature of dark matter (DM), the electroweak hierarchy problem, and the fermion mass hierarchy. 
Some of these puzzles can be addressed within the framework of strongly coupled gauge theories where the SM Higgs boson emerges as a composite state or pseudo-Nambu-Goldstone boson (pNGB), see for a review~\cite{Contino:2010rs,Panico:2015jxa}. Regarding the DM candidate within strongly coupled theories, one possibility is that it is a pNGB state which ensures it is naturally lighter than the composite scale. Hence it forms a good weakly interacting massive particle (WIMP) DM candidate~\cite{Frigerio:2012uc,Balkin:2017aep,DaRold:2019ccj,Ahmed:2020hiw} where DM is produced through the standard freeze-out mechanism assuming it remains in thermal equilibrium at temperatures larger than DM mass. There is a vast experimental program of DM searches including direct detection, indirect detection, and collider searches for a WIMP DM, see e.g.~\cite{Arcadi:2017kky}. However, there is no signal of WIMP DM to date, which motivates us to explore DM candidates beyond the WIMP or in general DM production through the freeze-out paradigm. 

In this work, we consider a framework where the SM and DM belong to a (strongly coupled) conformal/scale-invariant theory. We assume the conformal invariance is spontaneously broken at scale $f\gg v_{\rm SM}$, where $v_{\rm SM}\!=\!246\gev$ is the electroweak scale. 
Furthermore, we assume that the SM and DM do not have direct interactions below the conformal invariance breaking scale $f$. 
The spontaneous breaking of scale invariance results in a pseudo-Goldstone boson called, \emph{dilaton}~\cite{Salam:1969bwb}.
Requiring the scale invariance of the theory dictates the form of dilaton interactions with the SM and DM in the low energy effective theory below scale $\Lambda=4\pi f$~\cite{Goldberger:2008zz,Csaki:2007ns,Chacko:2012sy,Bellazzini:2012vz,Ahmed:2019kgl}.
It turns out that the dilaton interacts with the SM and DM through dimension-five or higher operators suppressed by the scale $f$. 
Hence the effective couplings between the SM and DM through the \emph{dilaton portal} is at least dimension-six order. 
For low-scale conformal breaking $f$, the SM and DM are in thermal equilibrium, and hence the production of DM via the dilaton portal is realized through the thermal freeze-out mechanism~\cite{Bai:2009ms,Lee:2013bua,Blum:2014jca,Efrati:2014aea,Kim:2016jbz,Ahmed:2019csf,Fuks:2020tam}.
However, assuming the conformal invariance breaking scale $f\gg v_{\rm SM}$, it is natural that the SM and DM are out of thermal equilibrium. Hence the possible DM production follows through the freeze-in mechanism, where DM is produced through the SM annihilation via the dilaton portal or through direct decays/annihilation of the dilaton field, see also~\cite{Brax:2021gpe,Brax:2021qfn,Kaneta:2021pyx,Barman:2021qds,Baldes:2021aph}.  

The freeze-in mechanism relies on the fact that initial DM abundance is negligible compared to the states in thermal equilibrium~\cite{Hall:2009bx}, see for a review~\cite{Bernal:2017kxu}. 
Furthermore, if the SM-DM interaction is a higher dimensional operator, as in our case, the freeze-in mechanism also crucially depends on the maximum temperature in the early universe. In particular, the DM production is dominated around the maximum temperature~\cite{Elahi:2014fsa}. The maximum temperature is usually assumed as the temperature at the end of reheating, $T_{\rm rh}$. This is the only correct description if the reheating is instantaneous. However, the reheating through perturbative decays of the inflaton field to the SM would be noninstantaneous, which leads to an extended period of reheating. Assuming a noninstantaneous reheating scenario, the maximum temperature $T_{\rm max}$ can be much larger than the temperature at the end of reheating phase $T_{\rm rh}$~\cite{Giudice:2000ex}. The maximum temperature $T_{\rm max}$ depends on inflaton energy density at the end of inflation, the equation of state $w$ during the reheating period, and the duration of reheating phase which can be parametrized by the temperature at the end of reheating phase, $T_{\rm rh}$. The end of reheating is defined when the SM energy density becomes equal to the inflaton energy density. Without specifying the details of the reheating phase, we consider the general equation of state $w$ in the range $(-1/3,1)$, where $w=0$ corresponds to the matter-dominated phase during the reheating process. 
Recently there have been significant studies on the implications of the reheating dynamics on the production of DM, see e.g.~\cite{Harigaya:2014waa,Chen:2017kvz,Kolb:2017jvz,Biswas:2019iqm,Ahmed:2019mjo,Chianese:2020yjo,Bernal:2020bfj,Garcia:2020eof,Bernal:2020qyu,Ahmed:2020fhc,Drees:2021lbm,Barman:2020plp,Mambrini:2021zpp,Barman:2021tgt,Garcia:2021iag,Ahmed:2021fvt,Ahmed:2022qeh,Ahmed:2022tfm}. 

We study the implications of nonstandard reheating dynamics on the ultraviolet (UV) freeze-in production of DM through the dilaton portal. We consider a DM candidate to be a scalar, fermion, or vector field. For simplicity, we assume no self-interactions for the DM field, and dark matter mass is the only parameter of the dark sector. 
The dilaton portal dynamics are completely fixed by two parameters; the dilaton mass and the conformal breaking scale $f$. 
Furthermore, the reheating dynamics is parametrized effectively by three parameters, the Hubble scale at the end of inflation, the equation of state $w$, and the reheating temperature. 
We study the freeze-in production of the DM within this framework and identify the parameter space where the observed DM abundance can be produced.

The paper is organized as follows: In \sec{sec:model} we present details of the dilaton portal DM model with the effective low energy Lagrangian including interactions of the dilaton field with the SM and DM. In \sec{sec:cosmology}, we describe early universe cosmology with the nonstandard period of reheating defined with the equation of state $w$. Production of DM via UV freeze-in is given in \sec{sec:dark_matter}, where we consider DM production through SM annihilation as well as dilaton annihilation/decays when kinematically allowed. Finally, in \sec{sec:conc} we conclude our findings. 

\section{The model}
\label{sec:model}
In this section, we lay out a framework where freeze-in dark matter production via a dilaton portal is realized. We assume a UV completion of the SM and DM involving a strongly coupled nearly scale-invariant theory. The scale invariance is broken spontaneously and the corresponding pseudo-Goldstone boson is the dilaton $\sigma(x)$. 
Furthermore, we assume the scale symmetry is realized nonlinearly below the symmetry breaking scale $f$, such that under the scale transformation $x^\mu\to x^{\prime\mu}\!=\!e^{-\omega} x^\mu $,  the dilaton undergoes a shift symmetry $\sigma(x)\to\sigma^\p(x^\p)\!=\!\sigma(x)+\omega f$. 
It is instructive to express the dilaton field as a {\it conformal compensator}, i.e.
\beq
\chi(x)=f e^{\sigma(x)/f},
\eeq
such that it transforms linearly under the scale transformation, i.e. $\chi(x)\to\chi^\p(x^\p)\!=\!e^{\omega}\chi(x)$. 
The vacuum expectation value (VEV) of $\chi(x)$ sets the scale of spontaneous symmetry breaking, i.e. $\langle \chi(x)\rangle\!\equiv\!f$, which is determined by the underlying strong sector dynamics at $\Lambda\!=\!4\pi f$. 

In this work, we consider the interactions of DM with the SM only through the dilaton portal.
In particular, we consider the following form of the Lagrangian
\begin{align}
{\cal L}&= {\cal L}_{\rm SM}+{\cal L}_{\rm DM}+{\cal L}_{\rm dil}+{\cal L}_{\rm SM}^{\rm int}+{\cal L}_{\rm DM}^{\rm int},		\label{eq:lag}
\end{align}
where ${\cal L}_{\rm SM}$ is the SM Lagrangian. As mentioned in the Introduction, we consider three possibilities for DM $X$, i.e. scalar DM (SDM), fermion DM (FDM), or vector DM (VDM), with the following Lagrangian,
\begin{align}
{\cal L}_{\rm DM}&=\begin{dcases}\frac{1}{2}\partial_\mu X\partial^\mu X-\frac{1}{2}m_{\!X}^2 X^2, &\text{SDM}\\
i\overline{X}\slashed{\partial} X-m_{\!X} \overline{X} X, &\text{FDM}\\
\!\!-\frac{1}{4}X_{\mu\nu}X^{\mu\nu}+\frac{1}{2}m_{\!X}^2 X_\mu^2, &\text{VDM}
\end{dcases}	\label{eq:lag_dm}
\end{align}
where $X_{\mu\nu}\!=\!\partial_\mu X_\nu-\partial_\nu X_\mu$ is the field strength tensor to the vector DM $X_\mu$.
The dilaton Lagrangian is
\begin{align}
\mathcal{L}_{\rm dil}=\frac{1}{2} \partial_{\mu} \chi \partial^{\mu} \chi-V(\chi),		\label{eq:lag_dilchi}
\end{align}
where we assume the following form for the dilaton potential,
\begin{align}
V(\chi)&=\frac{m_\sigma^2}{4f^2}\, \chi^{4} \bigg[\ln \bigg(\frac{\chi}{f}\bigg)-\frac{1}{4}\bigg],	\label{eq:pot_dil}
\end{align}
where we neglected terms of the order $O(m_\sigma^4/f^4)$ and higher. 

It is instructive to understand the origin of the above effective dilaton potential. In our framework, we employ a strongly coupled CFT theory which is explicitly broken at scale $\Lambda=4\pi f$ by a deformation operator $\op_{\!\rm def}(x)$ with scaling dimension $\Delta_{\rm def}$, such that 
\beq
{\cal L}_{\rm UV}\supset {\cal L}_{\rm CFT}+\lambda_{\rm def} \op_{\!\rm def}(x), 		\label{eq:L_UV}
\eeq
where parameter $\lambda_{\rm def}$ has dimension $\epsilon\equiv 4-\Delta_{\rm def}$. The deformation operator $\op_{\!\rm def}(x)$ transforms under the scale transformation as $\op_{\!\rm def}(x)\to\op_{\!\rm def}^\p(x^\p)\!=\!e^{\omega \Delta_{\rm def}}\op_{\!\rm def}(x)$. One can obtain the above effective dilaton potential~\eqref{eq:pot_dil} by either a spurion analysis~\cite{Goldberger:2008zz} or by a general (with one or higher loop) analysis~\cite{Chacko:2012sy} in this theory. Here we present the spurion analysis such that the spurion field corresponding to the CFT deformation allows the following form of nonderivative interactions for the dilaton potential~\cite{Goldberger:2008zz},
\beq
V(\chi)=\chi^4\sum_{j=0}^{\infty} c_j(\Delta_{\rm def}) \bigg(\frac{\chi}{f}\bigg)^{j(\Delta_{\rm def}-4)}\,,
\eeq
where $c_j\sim \lambda^j_{\rm def}$ are order one coefficients which in general depend on the scaling dimension and CFT breaking dynamics. Since we are interested in strongly coupled CFT, therefore the parameter $\lambda_{\rm def}$ cannot be arbitrarily small. Hence, treating the CFT deformation as small is possible only if the deformation operator $\op_{\!\rm def}(x)$ is nearly marginal, i.e. $\epsilon\equiv|4-\Delta_{\rm def}|\ll 1$. In this case, the above effective potential can be calculated with small order parameter $\epsilon$ as~\cite{Goldberger:2008zz} (see also~\cite{Chacko:2012sy})
\beq
V(\chi)=\frac{\epsilon}{4}\, \chi^{4} \bigg[\ln \bigg(\frac{\chi}{f}\bigg)-\frac{1}{4}\bigg]+O(\epsilon^2),
\eeq
where $\epsilon\ll1$.
Note that in \eq{eq:pot_dil} we have identified $m_\sigma^2=\epsilon \,f^2$, such that for $\epsilon\ll 1$ the dilaton mass $m_\sigma$ can be much smaller than the conformal breaking scale. In this work, we treat the dilaton mass $m_\sigma$ (in other words~$\epsilon$) as a free parameter. 

Before moving forward, we would like to comment on a possible UV completion of our model. We assume a strongly coupled CFT which in the low energy results in a dilaton field that interacts with the SM and DM as described in effective Lagrangian~\eqref{eq:lag}. This framework can be naturally realized in a holographic model with RS-like 5D warped extra dimension~\cite{Randall:1999ee} which involves two branes associated with a UV scale~$M_{\rm UV}$ and an IR scale~$\Lambda$. With the Goldberger-Wise~\cite{Goldberger:1999uk} stabilization mechanism of the radius of 5D warped extra dimension, one can identify the corresponding radion field as the dilaton of the 4D boundary theory using holographic dictionary~\cite{Arkani-Hamed:2000ijo,Rattazzi:2000hs}. In this framework, the effective theory only involves the IR scale $\Lambda$, which is identified in the 4D theory as the CFT breaking scale. Unlike RS model~\cite{Randall:1999ee}, we consider $\Lambda\gg O(1) \tev$. Moreover, in our framework we assume the SM and dark sector as elementary fields, therefore in the holographic model they are localized on the UV brane. A comprehensive analysis of this holographic model is beyond the scope of the present work where we only consider an effective theory valid below the IR scale $\Lambda=4\pi f$.

It is convenient to rewrite the dilaton field $\chi$ in terms of canonically normalized physical dilaton fluctuation $\sigma$ by expanding the $\chi$ field around its VEV, i.e. $\chi=f+\sigma$.
Such that the dilaton Lagrangian in the canonical basis takes the form,
\begin{align}
\mathcal{L}_{\rm dil}\!=\!\frac{1}{2} \partial_{\mu} \sigma \partial^{\mu} \sigma\!-\!\frac{1}{2}m_{\sigma}^2\sigma^2\!-\!\frac{5}{3!} \frac{m_{\sigma}^2}{f} \sigma^3\!-\!\frac{11}{4!} \frac{m_{\sigma}^2}{f^2}\sigma ^4,		\label{eq:lag_dil}
\end{align}
where we neglect higher order $O(\sigma^5)$ terms.
The interactions of the dilaton field $\sigma$ with the SM and DM are dictated by the nonlinearly realized scale invariance below scale $\Lambda\!=\!4\pi f$. 

In this work we assume that scale invariance breaking scale $\Lambda$ is much larger than the electroweak scale, i.e. $\Lambda\gg 1\tev$. 
Furthermore, as we are interested in UV freeze-in production of dark matter, therefore usually temperatures involved are much larger than the electroweak scale. Taking this into account, we define dilaton interactions with the SM only in the electroweak symmetric phase (EWSP) for temperatures $T$ above critical temperature $T_c\sim150\gev$. 
Note the SM electroweak symmetry is restored at $T>T_c$ and taking into account the thermal corrections at the one-loop level, Higgs effective potential takes the following form, 
\beq
V(H,T)\simeq \mu^2(T)|H|^2+\lambda(T) |H|^4,
\eeq
where the Higgs quartic coupling is $\lambda(T)\!\sim\! \lambda\simeq0.13$, and the Higgs effective mass parameter $\mu^2(T)$ can be approximated as,  
\beq
\mu^2(T)\approx \begin{cases}
-\lambda\, v_{\rm EW}^2& \lsp T\lesssim T_c\,,\\
~\beta\, T^2 &  \lsp T> T_c\,,
\end{cases}
\eeq
with the parameter $\beta\sim0.4$~\cite{Quiros:1999jp}.

\subsection{SM--dilaton interactions}
All the SM gauge and fermion fields are massless in the electroweak symmetric phase. However, the SM Higgs doublet (four real scalar components, $h_i$ with $i=0,1,2,3$) mass is $m_h^2=\beta\, T^2$. Therefore in the SM sources of explicit breaking of scale invariance are the Higgs mass term and the renormalization scale through RGE running of the coupling constants. To make the Higgs mass term scale invariant one needs to rescale it with the conformal compensator as~\cite{Goldberger:2008zz,Chacko:2012sy}, 
\beq
m_h^2 |H|^2\to  m_h^2\frac{\chi^2}{f^2} |H|^2.
\eeq
One can also think of this as the Higgs field being rescaled as $H\to H\chi/f$, since the Higgs VEV $\langle H\rangle$ explicitly breaks the scale invariance. 
In this case dilaton field $\sigma$ coupling with the Higgs field is
\begin{align}
{\cal L}_{\rm SM}^{\rm int}&\supset \sum_{i=0}^3\bigg(\frac{\sigma}{f}+\frac{\sigma^2}{2f^2}\bigg)\bigg[\partial_{\mu} h_i \partial^{\mu} h_i-2 m_{h}^{2} h_i^{2}\bigg],		\label{eq:lag_dilaton_higgs}
\end{align}
where higher-order interaction terms are neglected. Furthermore, the dilaton interacts with the SM fermions through the Yukawa terms, 
\begin{align}
{\cal L}_{\rm SM}^{\rm int}&\supset \frac{\sigma}{f}\bigg[y_t\bar Q_L \widetilde H t_R +\ldots\bigg]+\hc,		\label{eq:lag_dilaton_yukawa}
\end{align}
where $\widetilde H=i\sigma_2 H^\ast$ and ellipsis represent the fermions with smaller Yukawa couplings. Since top Yukawa coupling $y_t\sim 1$ is the largest coupling, it would be the most relevant for our analysis. 
The dilaton interactions with the SM massless gauge bosons emerge due to RGE running of their gauge couplings as,
\begin{align}
{\cal L}_{\rm SM}^{\rm int}&\supset \sum_{i=1}^3 \frac{\alpha_i}{8 \pi} b_i  \frac{\sigma}{f}F_{i\mu \nu} F^{\mu \nu}_i,	\label{eq:lag_dilaton_gauge}
\end{align}
where $i=1,2,3$ corresponds to SM gauge groups $U(1)_Y$, $SU(2)_L$, and $SU(3)_c$, respectively. Whereas $\alpha_i=g_i^2/4\pi$ and $b_i$ are the corresponding gauge couplings and beta-function coefficients, respectively.

\subsection{DM--dilaton interactions}
At the leading order, the interaction Lagrangian for the dilaton field with the DM is given by, 
\begin{align}
{\cal L}_{\rm DM}^{\rm int}&\!=\!\begin{dcases}\!\!\bigg(\!\frac{\sigma}{f}\!+\!\frac{\sigma^2}{2f^2}\!\bigg)\!\!\bigg[\!\partial_{\mu} X\partial^{\mu} X\!-\!2 m_{X}^{2} X^{2}\!\bigg], &\text{SDM}\\
\!\!-\frac{\sigma}{f}m_{\!X} \overline{X} X, &\text{FDM}\\
\!\!\bigg(\frac{\sigma}{f}+\frac{\sigma^2}{2f^2}\bigg)m_{\!X}^2 X_\mu^2,	\qquad&\text{VDM}
\end{dcases}	\label{eq:lag_dm_int}
\end{align}
where we consider fermion DM as a Dirac particle, however for Majorana fermion the above interaction term is rescaled by factor 1/2. 

Note that the above interactions of the dilaton field with the elementary SM or DM fields can be obtained through the argument that the dilaton $\chi$ couples to the trace of energy-momentum tensor as~\cite{Goldberger:2008zz}
\begin{align}
&{\cal L}_{\rm SM\!/\!DM}^{\rm int}= \frac{\chi}{f} \,T_{\mu,{\rm SM/DM}}^\mu,	\\
&=\frac{\chi}{f}\bigg[\sum_i g_i(\mu)(\Delta_i-4)\op^i_{\!\rm SM\!/\!DM}	\notag\\
&\qquad+\sum_i\beta_i(g_i)\frac{\partial {\cal L}_{\rm SM\!/\!DM}}{\partial g_i}\bigg], \notag
\end{align}
where $\Delta_i$ is the scaling dimension of the SM/DM operator $\op^i_{\!\rm SM\!/\!DM}$ and $\beta_i\equiv \mu\, \partial g_i/\partial \mu$ is the beta-function.

\section{Nonstandard cosmology during reheating}
\label{sec:cosmology}
We assume a slow-roll inflationary paradigm with quasi-de Sitter expansion of the Universe with Hubble parameter $H_{\! I}$. 
The inflationary epoch of accelerated expansion ends with a reheating phase where the inflaton field $\phi$ is assumed to transfer its energy density via perturbative decays to the SM sector. 
For concreteness we consider $\alpha$-attractor T-model of inflation~\cite{Kallosh:2013hoa,Kallosh:2013yoa} with inflaton potential
\begin{align}
V(\phi) &=\Lambda_{I}^{4} \tanh ^{2 n}\bigg(\!\frac{|\phi|}{M}\!\bigg)	\notag\\
&\simeq \begin{dcases}
\Lambda_{I}^{4}, & |\phi| \gg\! M\\
\Lambda_{I}^{4}\Big\vert\frac{\phi}{M}\Big\vert^{2 n},  & |\phi| \ll\! M
\end{dcases},\label{eq:inf_pot}
\end{align} 
where $\Lambda_{I}$ determines the scale of inflation, whereas $M$ is related to the reduced Planck mass through the $\alpha$ parameter as $M\equiv\sqrt{6 \alpha}\, \mpl$. The above potential $V(\phi)$ approximates to constant value for $|\phi| \!\gg\! M$ which is ideal for slow-roll inflation. Whereas, the inflaton potential takes a monomial form proportional to $|\phi|^{2 n}$ for $|\phi|\! \ll\! M$. In this model the inflationary phase ends when $|\phi| \!\sim\! M$ and for smaller field values the inflaton field coherently oscillates around its minimum at $\phi\!=\!0$ for positive values of~$n$. In this regime, we assume the inflaton field perturbatively decays to the SM sector, which is referred to as the reheating phase. 

Without specifying the details of reheating dynamics, we assume inflaton energy density scales as $\rho_\phi\propto a^{-3(1+w)}$ during the reheating phase, where $w$ is the equation of state. 
During inflation $w\simeq -1$ and the end of inflation is marked when $w=-1/3$. 
During the reheating phase total energy density is dominated by the inflaton energy density where the equation of state parameter $w$ can be related to the inflaton potential parameter $n$ as,
\beq
w=\frac{\langle p_\phi\rangle}{\langle \rho_\phi\rangle}=\frac{n-1}{n+1}.	\label{eq:w}
\eeq
Above $\langle p_\phi\rangle$ and $\langle \rho_\phi\rangle$ are the inflaton pressure and energy density which are time-averaged over one inflaton oscillation.
In the following, we assume the equation of state $w\in(-1/3,1)$ during the reheating phase. For example, $w=0$ defines the matter-dominated universe due to the inflaton coherent oscillations. The Hubble rate is given by, 
\beq
H(a)\equiv \frac{\dot a}{a}= \sqrt{\frac{\rho (a)}{3\mpl^2}},
\eeq
where $\mpl=2.435\!\times\! 10^{18}\gev$ is the reduced Planck mass and {\it over-dot} is derivative with respect to time $t$. Above the energy density $\rho (a)$ is the sum of the inflaton and the SM radiation energy densities, i.e. $\rho (a)=\rho_\phi(a)+\rho_{\!R}^{}(a)$. During the reheating phase inflaton energy density is dominant, i.e. $\rho_\phi(a)\gg\rho_{\!R}^{}(a)$ for $a_e<a<a_{\rm rh}$, where $a_e$ and $a_{\rm rh}$ denote the end of inflation and the reheating phase, respectively. Furthermore, we assume that DM energy density $\rho_{\!X}^{}$ remains a subdominant component of the total energy density. 
The end of the reheating phase is defined when $\rho_\phi(a_{\rm rh})=\rho_{\!R}^{}(a_{\rm rh})$.

The exact cosmological evolution of the reheating phase is determined by solving the coupled Boltzmann equations, 
\beq
\begin{aligned}
\dot{\rho}_{\phi}+3(1+w) H \rho_{\phi}&=-\Gamma_{\phi}\, \rho_{\phi},   	\\
\dot \rho_{\!R}^{}+4 H\rho_{\!R}^{}&=+\Gamma_{\phi}\, \rho_{\phi},
\end{aligned}
\eeq
where $\Gamma_\phi$ is the perturbative decay width of the inflation field to SM radiation. 
For instance, this can be achieved by employing an effective coupling between the inflaton and SM fermions $\psi$ of the form $y\phi \bar \psi \psi$, then the perturbative inflaton decay rate $\Gamma_\phi$ is, 
\beq
\Gamma_{\phi}=\frac{y^2}{8\pi}m_\phi. 
\eeq
Other possible inflaton decay channels to SM fields can also be considered. In general, $\Gamma_\phi$ is not a fixed quantity, but depends on time, see e.g.~\cite{Ahmed:2021fvt}. More exact computation of the right-hand-side of the above Boltzmann equations, i.e. $\Gamma_\phi \rho_\phi$, is more complicated as recently discussed in~\cite{Garcia:2020eof,Ahmed:2022tfm}. However, in this work, we remain agnostic about details of the reheating dynamics and for simplicity treat $\Gamma_{\phi}$ as a free parameter with a constant value.

Assuming $\Gamma_\phi\ll 3(1+w)H$ during the reheating phase $a_e\lesssim a\lesssim a_{\rm rh}$ we can approximately solve the above Boltzmann equations as, 
\begin{align}
\rho_{\phi}&\simeq3 \mpl^{2} H_{\!I}^{2}\bigg(\frac{a_{e}}{a}\bigg)^{3(1+w)}, 		\\
\rho_{\!R}^{}&\simeq 3 \mpl^{2} H_{\!I} H_{\rm rh}\bigg[\bigg(\frac{a_{e}}{a}\bigg)^{3(1+w)/2}-\bigg(\frac{a_{e}}{a}\bigg)^{4}\bigg],	\label{eq:rho}
\end{align}
where we employed $\Gamma_\phi\!\simeq\! (5-3w)H_{\rm rh}/2$ with $H_{\rm rh}$ being the Hubble scale at the end of reheating,  
\beq
H_{\rm rh}\equiv H(a_{\rm rh}) = H_{\!I}\Big(\frac{a_{e}}{a_{\mathrm{rh}}}\Big)^{3(1+w)/2}\,. 	\label{eq:Hrh}
\eeq
Above the Hubble scale during inflation $H_{\!I}$ is related to the inflaton potential during inflation, i.e. $\phi\!>\!\mpl$, as
\beq
H_{\!I}^2=\frac{\rho_\phi}{3\mpl^2}\simeq\frac{V(\phi)}{3\mpl^2}\simeq\frac{\Lambda^4}{3\mpl^2}.		\label{eq:HI}
\eeq
For the slow-roll inflationary scenario the recent Planck~\cite{Planck:2018jri} and {\it BICEP/Keck}~\cite{BICEP:2021xfz} measurements put an upper bound at 95\% C.L. on the Hubble scale during inflation as
\beq
H_{\! I}\lesssim 4.4\times10^{13}\gev.
\eeq
Hence, the above result implies an upper limit on the inflationary scale as $\Lambda_{I} \lesssim1.4\times 10^{16}\gev$. Furthermore, the current upper bound on the tensor to scalar power spectrum ratio, $r\lesssim 0.036$~\cite{BICEP:2021xfz} at 95\% C.L., limits the value of the $\alpha$ parameter or $M$ from above, such that, \mbox{$M\!\lesssim\! 10 \mpl$}.  
Hereinafter, without loss of generality, we fix $\alpha\!=\! 1/6$, such that $M\!=\!\mpl$. The scale of inflation $\Lambda_I$ is related to the Hubble scale during inflation $H_{\! I}$ through \eq{eq:HI}, which is the only free parameter during the inflationary phase. Whereas, during the reheating phase the inflaton potential parameter~$n$ is related to the equation of state $w$ via~\eq{eq:w}, which we treat as a free parameter with values $w\in(-1/3,1)$.  

After the end of reheating, the inflaton energy density rapidly vanishes and standard cosmological evolution takes its course where SM radiation is the dominant energy density until the matter-radiation equality, i.e. $a\!=\!a_{\rm eq}$. During the radiation-dominated epoch, $a_{\rm rh}\lesssim a\lesssim a_{\rm eq}$, the radiation energy density is given as 
\beq
\rho_{\!R}^{}\simeq \rho_{\rm rh}\Big(\frac{a_{\mathrm{rh}}}{a}\Big)^{\!4}, 	\qquad {\rm where}\quad \rho_{\rm rh}\equiv 3 \mpl^{2} H_{\mathrm{rh}}^{2}. 
\eeq
It is instructive to write the Hubble parameter as a function of the scale factor 
\begin{align}
H(a) &= \begin{dcases}
  H_{\! I} \Big(\frac{a_e}{a}\Big)^{\frac{3 (1+w)}{2}}, & a_e < a \leq a_{\rm rh}\,,\\
  H_{\rm rh} \Big(\frac{a_{\rm rh}}{a}\Big)^{2}, 	\qquad&  a_{\rm rh}<a\leq a_{\rm eq}\,,
  \end{dcases}	\label{eq:Ha}
\end{align}
where $H_{\rm rh}$ is defined in \eqref{eq:Hrh}.
The temperature of the SM bath is defined in terms of the SM radiation energy density as
\begin{align}
T^4&\!=\!\frac{30\,\rho_{\!R}^{}}{\pi^2 g_\star(T)}\,,		\notag \\
&\!\simeq\! \frac{g_\star(T_{\rm rh})}{g_\star(T)} T_{\rm rh}^4 \begin{dcases}
  \frac{H(a)}{H_{\rm rh}}, & a_{\rm max} \!<\! a \!\leq\! a_{\rm rh}\,,\\
  \!\!\Big(\frac{H(a)}{H_{\rm rh}}\Big)^{\!\!2}, 	&  a\!>\!a_{\rm rh}\,, 
 \end{dcases}	\label{eq:Ta}
\end{align}
where $a_{\rm max}$ is the value of scale factor when the temperature (or radiation energy density) has its maximum value, see below \eq{eq:Tmax}. Whereas, $g_\star(T)$ is the effective number of relativistic d.o.f. contributing to the energy density. Temperature at the end of the reheating period $T_{\rm rh}$ is defined as
\beq
T_{\rm rh}^2\equiv\frac{3}{\pi}\sqrt{\frac{10}{g_\star(T_{\rm rh})}}\,\mpl H_{\rm rh}.	\label{eq:Trh}
\eeq 
In the following analysis, we treat temperature $T_{\rm rh}$ as one of the free parameters of the model. 

Note that the initial condition for the radiation energy density at $a=a_e$ is $\rho_{\!R}^{}(a_e)=0$, therefore the temperature of the SM bath is also zero at the onset of the reheating phase. The maximum of radiation energy density or the maximum temperature $T_{\rm max}$ is reached during the reheating phase at $a=a_{\rm max}<a_{\rm rh}$, 
\beq
a_{\rm max}=a_e\,\bigg(\frac{8}{3 (1+w)}\bigg)^{\!\frac{2}{5-3 w}}, 		\quad {\rm for}~-\!\frac{1}{3}\!<\!w\!<\!\frac{5}{3}. 
\eeq 
Hence using \eqref{eq:rho} and \eqref{eq:Trh} the corresponding maximum temperature $T_{\rm max}$ can be written as, 
\beq
T_{\rm max}^4\!=\!\frac{3(5-3 w)}{8\pi}\!\bigg(\frac{3 (1+w)}{8}\bigg)^{\!\!\frac{3(1+ w)}{5-3 w}}\!\sqrt{\frac{10}{\!g_\star(T_{\rm rh})\!}}\, \mpl H_{\!I} \, T_{\rm rh}^2,	\label{eq:Tmax}
\eeq
for $-1/3<w<5/3$.
In the following analysis, we have three free parameters which determine nonstandard cosmological evolution during the reheating phase, namely (i) the Hubble scale at the end of inflation $H_{\! I}$, (ii) the temperature at the end of reheating $T_{\rm rh}$, and (iii) the equation of state $w$ which we take in the range $(-1/3,1)$. The value of the scale factor at the end of inflation $a_e$ is arbitrary. With these three free parameters $(H_{\! I}, T_{\rm rh},w)$ one can readily calculate all the other quantities related to cosmological and thermal evolution. Now the inflationary Hubble scale is constrained by the CMB measurement of the inflationary perturbations. The current upper bound from Planck data~\cite{Planck:2018vyg} reads as $H_{\! I}\lesssim 6\!\times\!10^{13}\gev$. An upper bound on the maximum temperature $T_{\rm max}$ is set by the requirement that radiation energy density is smaller than the total energy density at the end of inflation, i.e. $\rho_{\!R}^{}\lesssim 3\mpl^2 H_{\!I}^2$. Employing the upper bound on $H_{\!I}$, we get $T_{\rm max}\lesssim 6.5\!\times\!10^{15}\gev$.
Reheating temperature is by default smaller than the maximum temperature, i.e. $T_{\rm rh}< T_{\rm max}$. Furthermore, the BBN sets a lower bound on the reheating temperature $T_{\rm rh}\gtrsim 1\mev$~\cite{Sarkar:1995dd}.

\section{Dilaton portal dark matter production}
\label{sec:dark_matter}
In this section, we discuss the freeze-in production of dark matter via the dilaton portal. The Boltzmann equation for DM $X$ is, 
\beq
\dot n_{\!X} +3 H n_{\!X}={\cal C}_{X}+{\cal D}_{X},	\label{e.boltzmann}
\eeq
where ${\cal C}/{\cal D}$ are the collision/decay terms,
\begin{align}
{\cal C}_{X}&\simeq \bar n_{\!X}^2 \Big[ \langle \sigma_{XX\to {\rm SM\, SM}}^{}\, v\rangle+ \langle \sigma_{XX\to \sigma\sigma}^{}\, v\rangle\Big],		\\
{\cal D}_{X}&\simeq 2 n_\sigma \langle \Gamma_{\sigma\to XX}\rangle,		\label{e.decay}
\end{align}
with $\langle \sigma_{XX\to {\rm SM\, SM}^{}}\, v\rangle$ and $\langle \Gamma_{\phi\to XX}\rangle$ being the thermally averaged annihilation cross section and partial width, respectively. 
Above the equilibrium number density of species $i$ with spin $J_i$ is defined as
\beq
\bar n_i = \left(2 J_{i}+1\right) \frac{m_{i}^{2}\, T}{2 \pi^{2}} K_{2}\!\Big(\frac{m_{i}}{T}\Big),
\eeq
where $K_2(x)$ is the Bessel function of the second kind. In the relativistic regime, i.e. $T\gg m_i$, the above $\bar n_i$ receives a correction due to quantum statistics of order $\zeta(3)\simeq 1.2$ and $3\zeta(3)/4\simeq 0.9$ for bosons and fermions, respectively. However, for simplicity, we neglect this correction in our analysis. 
Furthermore, we assume no interaction between the inflaton and dilaton, therefore dilaton is produced through scattering and inverse decays of the SM fields. 
Above $n_\sigma$ is the number density of the dilaton field. If the interaction rate between the SM and dilaton is larger than the Hubble scale then dilaton is in thermal equilibrium, i.e. $n_\sigma=\bar n_\sigma$. In general, we solve the following Boltzmann equation to get the dilaton number density $n_\sigma$, 
\beq
\dot n_{\sigma} +3 H n_{\sigma}=\bar n_\sigma\langle \Gamma_{{\rm SM}\to \sigma}\rangle, 
\eeq
where the thermally averaged SM annihilation to dilaton $\langle \Gamma_{{\rm SM}\to \sigma}\rangle$ is dominated by the top-Yukawa interaction term~\eq{eq:lag_dilaton_yukawa}. 

It is convenient to recast the Boltzmann equation~\eqref{e.boltzmann} in terms of the comoving number density $N_{\!X}=n_{\!X} a^3$ and as a function of temperature $T$,
\beq
\begin{aligned}
\frac{dN_{\!X}}{dT} &= -\frac{3a_{\rm rh}^3}{\pi} \sqrt{\frac{10}{g_\star(T)}} \mpl \frac{T_{\rm rh}^3}{T^6}\Big[{\cal C}_{X}+{\cal D}_{X}\Big]	\\
&\qquad\times\!
\begin{dcases}
\frac{8}{3(1+w)}\bigg(\frac{T_{\rm rh}}{T}\bigg)^{\frac{7-w}{1+w}}\,,      	& T \geq T_{\rm rh},\\
\frac{g_{\star s}(T_{\rm rh})}{g_{\star s}(T)}\,,	&T < T_{\rm rh},
\end{dcases}	\label{e.NX}
\end{aligned}
\eeq
where $g_{\star s}(T)$ denote the effective number of relativistic d.o.f. contributing to entropy density $s=2\pi^2 g_{\star s}(T)\, T^3/45$. Dark matter present relic abundance can be calculated as 
\beq
\Omega_{X} h^{2}=\frac{m_{X} n_{X}(a_{0})}{\rho_{c}\, h^{-2}}=\frac{m_{X}}{\rho_{c}\, h^{-2}}\frac{N_{X}(T_{0})}{a_{\rm rh}^3}\frac{s_0}{s_{\rm rh}}, 	\label{e.omegah2}
\eeq
where $n_{\!X}(a_0)$ is the DM number at present. Whereas, the second equality is obtained employing entropy conservation with present entropy density $s_0=2970\,{\rm cm^{-3}}$ and $s_{\rm rh}\equiv s(T_{\rm rh})$. Above the critical energy density is $\rho_{c}=1.054 \times 10^{-5} h^{2}\, {\rm GeV\, cm^{-3}}$

\subsection{SM and dilaton scattering}
\label{sec:sm_annihilation}

\begin{figure}[t!]
\centering
\includegraphics[width=\linewidth]{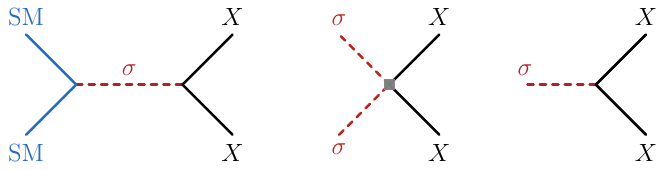}
\vspace{-7pt}
\caption{Schematic diagrams for the freeze-in production of DM through the dilaton portal.}
\label{fig:SM_DM_DP}
\end{figure}

The schematic diagrams contributing to the production of DM due to the annihilation of SM bath particles as well as the annihilation/decay of the dilaton field are shown in \fig{fig:SM_DM_DP}. 
Dark matter production through the annihilation of the dilaton field is relevant when the latter is in thermal equilibrium with the SM bath. As mentioned above, since we are interested in the production of DM through UV freeze-in, we consider SM annihilation to DM only in the electroweak symmetric phase. 

\subsubsection{Scalar DM case} 
For the case of vector DM $X_\mu$ amplitude squared for the SM and dilaton scattering processes are, 
\begin{align}
|{\cal M}_{XX\to hh}|^2&=\frac{\left(2 m_h^2+s\right)^2 \left(2 m_{\!X}^2+s\right)^2}{2 f^4 \left(m_{\sigma }^2 \left(\Gamma _{\sigma }^2-2 s\right)+m_{\sigma }^4+s^2\right)},	\\
|{\cal M}_{XX\to V_i V_i}|^2&=\frac{b_{i}^2 \alpha_{i}^2 }{4\pi^2}\frac{s^2 \left(2m_{\!X}^2+s\right)^2}{4f^4 \left(\Gamma _{\sigma }^2 m_{\sigma }^2+\left(s-m_{\sigma }^2\right)^2\right)}	,	\\
|{\cal M}_{XX\to \sigma\sigma}|^2&= \frac{1}{2 f^4}\bigg[\frac{\left(m_{\sigma }^2+2 m_{\!X}^2\right){}^2}{u-m_{\!X}^2} +\frac{\left(m_{\sigma }^2+2 m_{\!X}^2\right){}^2}{t-m_{\!X}^2}	\notag\\
&\hspace{-11pt}-\frac{2 m_{\sigma }^2 \left(m_{\sigma }^2+8 m_{\!X}^2+3 s\right)}{m_{\sigma }^2-s}-6 m_{\!X}^2\bigg]^2,
\end{align}
where $b_i$ and $\alpha_i$ are the beta-function coefficient and the gauge couplings with $i=1,2,3$ for the SM gauge groups $U(1)_Y,SU(2)_L$, and $SU(3)_c$, respectively. 
The scattering rate ${\cal C}_{\!X}$ can be approximated for the above processes as,
\begin{align}
{\cal C}_{X}\!\approx\!
\begin{cases}\!\!\big(5+\frac{b_{i}^2 \alpha_{i}^2 }{8\pi^2}\big)\frac{3 T^8}{4 \pi ^5 f^4}\,, & T\gg m_\sigma,m_{\!X}	\\
\!\!\big(4+\frac{b_{i}^2 \alpha_{i}^2 }{8\pi^2}\big)\frac{2880 T^{12}}{\pi ^5 f^4 m_\sigma^4}\,, & m_\sigma \gg T\gg m_{\!X},	\\ 
\!\!\big(4+\frac{b_{i}^2 \alpha_{i}^2 }{8\pi^2}\big)\frac{9 m_{\!X}^9 T^3}{8 \pi ^4 f^4 m_\sigma^4} \,e^{\frac{-2 m_{\!X}}{T}}\,, & m_\sigma\gg m_{\!X}	 \gg T
\end{cases}	\label{eq_CX_scalar}
\end{align}
where summation over all the SM gauge bosons $i$ is assumed. 

\subsubsection{Fermion DM case}
For the case of vector DM $X_\mu$ amplitude squared for the SM and dilaton scattering processes are, 
\begin{align}
|{\cal M}_{XX\to hh}|^2&\!=\!\frac{49 m_{\!X}^2 (2 m_h^2+s)^2 (s-4 m_{\!X}^2)}{4 f^4 \big(\Gamma _{\sigma }^2 m_{\sigma }^2+(m_{\sigma }^2-s)^2\big)},	\\
|{\cal M}_{XX\to V_i V_i}|^2&\!=\!\frac{b_{i}^2 \alpha_{i}^2}{4\pi^2}\frac{49 s^2 m_{\!X}^2 (s-4 m_{\!X}^2)}{8 f^4 \big(\Gamma _{\sigma }^2 m_{\sigma }^2+(m_{\sigma }^2-s)^2\big)}	,	\\
|{\cal M}_{XX\to \sigma\sigma}|^2&=\frac{49 m_{\!X}^2 (s-4 m_{\!X}^2) (15 m_{\sigma }^2+s)^2}{16 f^4 \big(\Gamma _{\sigma }^2 m_{\sigma }^2+(s-m_{\sigma }^2)^2\big)}.
\end{align}
The scattering rate ${\cal C}_{\!X}$ for the fermion DM can be approximated as
\begin{align}
{\cal C}_{X}\!\approx\!
\begin{cases}\!\!\big(4+\frac{b_{i}^2 \alpha_{i}^2 }{8\pi^2}\big)\frac{m_{\!X}^2 T^6}{64 \pi ^5 f^4}+\frac{25 m_{\!X}^2 m_\sigma^4 T^2}{4096 \sqrt{2} \pi ^4 f^4}\,, & T\gg m_\sigma,m_{\!X}	\\
\!\!\big(4+\frac{b_{i}^2 \alpha_{i}^2 }{8\pi^2}\big)\frac{18 m_{\!X}^2 T^{10}}{\pi ^5 f^4 m_\sigma^4}\,, & m_\sigma \gg T\gg m_{\!X},	\\	
\!\!\big(4+\frac{b_{i}^2 \alpha_{i}^2 }{8\pi^2}\big)\frac{75 m_{\!X}^4 T^4}{512 \pi ^4 f^4} \,e^{\frac{-2 m_{\!X}}{T}}\,, & m_\sigma\gg m_{\!X}	 \gg T.
\end{cases} \label{eq_CX_fermion}
\end{align}

\subsubsection{Vector DM case} 
For the case of vector DM $X_\mu$ amplitude squared for the SM and dilaton scattering processes are, 
\begin{align}
|\!{\cal M}_{XX\to hh}\!|^2&\!=\!\frac{(s^2\!-4 s m_{\!X}^2\!+\!12 m_{\!X}^4\!)(2 m_h^2\!+\!s)^2}{2 f^4 \big(\Gamma _{\sigma }^2 m_{\sigma }^2+(s-m_{\sigma }^2)^2\big)},	\\
|\!{\cal M}_{XX\to V_i V_i}\!|^2\!&=\!\frac{b_{i}^2 \alpha_{i}^2}{4\pi^2}\frac{s^2\left(s^2-4 s m_{\!X}^2+12 m_{\!X}^4\right)}{4 f^4 \big(\Gamma _{\sigma }^2 m_{\sigma }^2+(s-m_{\sigma }^2)^2\big)},	\\
|\!{\cal M}_{XX\to \sigma\sigma}\!|^2&\!=\! \frac{(s^2\!-\!4 s m_{\!X}^2\!+\!12 m_{\!X}^4\!) (7 m_{\sigma }^2\!+\!s)^2}{2 f^4 \big(\Gamma _{\sigma }^2 m_{\sigma }^2+(s-m_{\sigma }^2)^2\big)}.
\end{align}
In this case, neglecting the leading order phase space factor, the approximate form of scattering rate ${\cal C}_{\!X}$ is the same as that of the scalar DM case, i.e. \eq{eq_CX_scalar}.

\begin{figure*}[t!]
\centering
\includegraphics[width=0.43\textwidth]{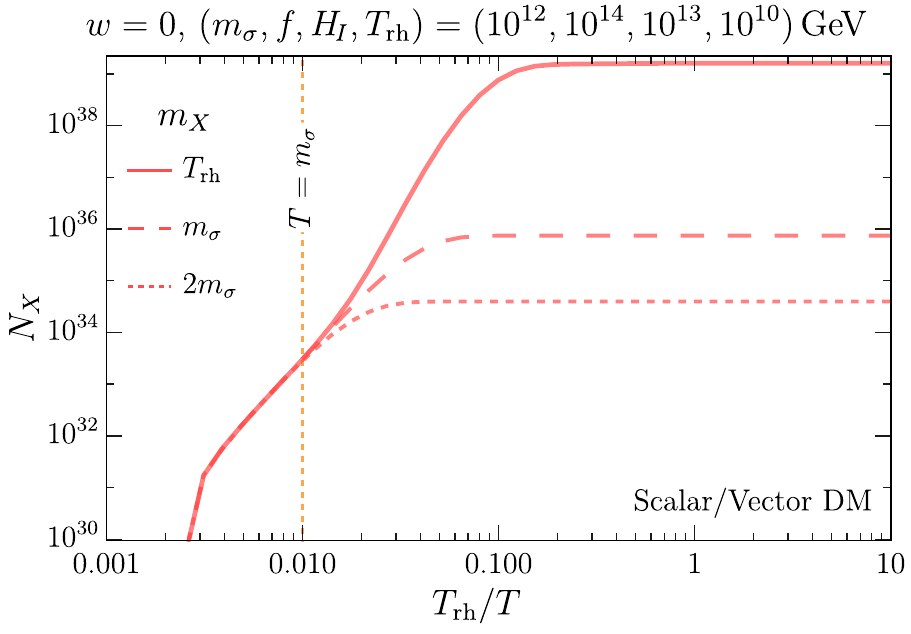}\!\!\!
\includegraphics[width=0.43\textwidth]{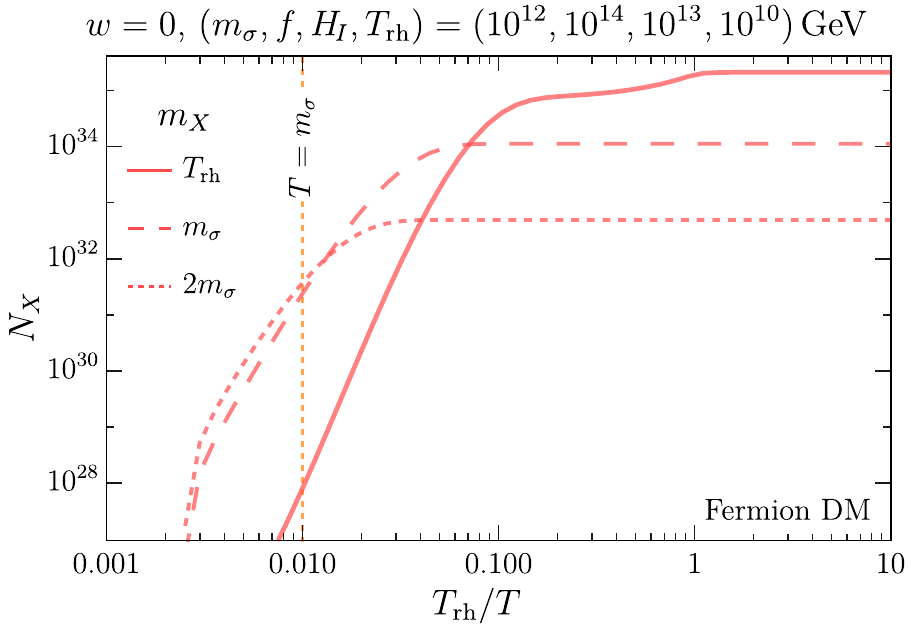}
\caption{Comoving number density $N_{\!X}$ for scalar/vector DM (left panel) and Dirac fermion DM (right panel) as a function of $T_{\rm rh}/T$ for different choices DM mass $m_{\!X}$. }
\label{fig:NT_mX}
\end{figure*}
\subsection{Dilaton decays}
\label{sec:dilaton_decay}
For dark matter lighter than dilaton, the dominant contribution to the production of DM is through dilaton decay. In this case, the decay term ${\cal D}_{\!X}$ in \eq{e.boltzmann} is given by 
\begin{align}
{\cal D}_{\!X}&=2n_\sigma\frac{K_1\big(\!\frac{m_\sigma}{T}\!\big)}{K_2\big(\!\frac{m_\sigma}{T}\!\big)} \\
&\qquad\times\!\begin{cases}\!\!\frac{m_{\sigma }^3}{32 \pi  f^2} \sqrt{1-\frac{4 m_{\!X}^2}{m_{\sigma }^2}} \left(1+\frac{2 m_{\!X}^2}{m_{\sigma }^2}\right)^2, &\text{SDM}\\
\!\!\frac{m_{\sigma }m_{\!X}^2}{8 \pi  f^2}  \left(1-\frac{4 m_{\!X}^2}{m_{\sigma }^2}\right)^{3/2}, &\text{FDM}\\
\!\!\frac{m_{\sigma }^3}{32 \pi  f^2} \sqrt{1-\frac{4 m_{\!X}^2}{m_{\sigma }^2}} \left(1-\frac{4 m_{\!X}^2}{m_{\sigma }^2}+\frac{12 m_{\!X}^4}{m_{\sigma }^4}\right),	&\text{VDM}
\end{cases}	\label{eq:dilaton_decay}	\notag
\end{align}
where $n_\sigma$ is the number density of the dilaton field. We note that in most natural parameter space when $f\gg v_{\rm SM}$ the dilaton field is out of thermal equilibrium with the SM bath, therefore the number density of the dilaton $n_\sigma$ is calculated through the SM annihilation and inverse decays of the SM fields. The dominant contribution to the dilaton production is due to scattering processes involving top-Yukawa coupling~\eqref{eq:lag_dilaton_yukawa}, i.e. $y_t Q_L \widetilde H t_R$, see also~\cite{Brax:2021gpe,Kaneta:2021pyx}. The condition for dilaton to be in thermal equilibrium reads as 
\beq
\begin{aligned}
1< \frac{\langle \Gamma_{{\rm SM}\to \sigma}\rangle}{H} &\sim  0.04\frac{y_t^2T^3}{f^2 H}  \\
&\sim  0.01\frac{\mpl }{f^2}
\begin{dcases} \frac{T_{\rm rh}^2}{T}, \quad& T\geq T_{\rm rh}\\
T,	& T< T_{\rm rh}
\end{dcases}	
\end{aligned}\label{eq:dilaton_decay}
\eeq
which maximizes for $T=T_{\rm rh}$ and the thermalization condition becomes $T_{\rm rh}\gtrsim 10^{2} f^2/\mpl$.

\subsection{Numerical analysis}
\label{sec:results}
Making use of the results from the previous section our goal is to solve the Boltzmann equation~\eqref{e.NX} to calculate the comoving number density and thus the relic abundance using \eq{e.omegah2}. Before discussing the numerical results, we observe that our model contains six free parameters: $m_{\!X}$, $m_\sigma$, $f$, $H_{\! I}$, $T_{\rm rh}$, and $w$. Note, however, that dilaton mass cannot be arbitrarily smaller than the scale invariance breaking scale $f$. Naturalness suggests the dilaton mass of the same order as breaking scale $f$, therefore $m_\sigma/f$ shows the amount of tuning. In the following, we show the dependence of dark matter comoving number density w.r.t temperature evolution for various choices of these parameters. 

\subsubsection{Dependence on DM mass} 
In \fig{fig:NT_mX}, we present $N_{\!X}$ as a function of $T_{\rm rh}/T$ for different choices of DM mass for scalar/vector DM (left panel) and fermion DM (right panel). We fix the remaining parameters as: $w=0$, $f=10^{14}\gev$, $H_{\! I}=10^{13}\gev$ and $T_{\rm rh}=10^{10}\gev$, with $m_\sigma/f=1\%$. 
For this choice of parameters maximum temperature obtained is $T_{\rm max}\sim 10^{3} T_{\rm rh}$.
In this figure, we show $T=m_\sigma=10^{12}\gev$ with an orange dashed vertical line, to the left of this line $T>m_\sigma$ and hence dilaton can be produced on-shell via inverse decays of SM fields, and therefore its decays to DM (for $m_\sigma>2m_{\!X}$) are the dominant source of DM production. However, for temperatures smaller than dilaton mass $T<m_\sigma$, the dominant source of DM production is SM annihilation to DM through an effective dimension-8 operator, where s-channel dilaton is integrated out. In \fig{fig:NT_mX} (left panel) for scalar/vector DM illustrates these features, where comoving number density acquires maximum value for temperatures $m_\sigma >T > T_{\rm rh}$ for DM masses $m_{\!X}<T_{\rm max}$. We consider three illustrative values for $m_{\!X}=T_{\rm rh}, m_\sigma, 2m_\sigma$, however, we note that for DM mass $m_{\!X}<T_{\rm rh}$, the results are same as that of the $m_{\!X}=T_{\rm rh}$ case.

The right panel of \fig{fig:NT_mX} shows the case when DM is a Dirac fermion. We consider the three values of DM mass $m_{\!X}=T_{\rm rh}, m_\sigma, 2m_\sigma$ shown as solid, long-dashed, and dashed curves, respectively. 
In this case, we also note that maximum comoving number density is attained when $m_\sigma >T > T_{\rm rh}$.
As mentioned above, the cross section is proportional to fermionic DM mass therefore for dark matter mass smaller than $T_{\rm rh}$ the DM number density scales as $m_{\!X}^2/T_{\rm rh}^2$ at the end of reheating. 
Note that once the DM mass becomes larger than the temperature, i.e. $m_{\!X}>T$, the number density becomes exponentially suppressed due to Boltzmann suppression. 
\begin{figure*}[t!]
\centering
\includegraphics[width=0.43\textwidth]{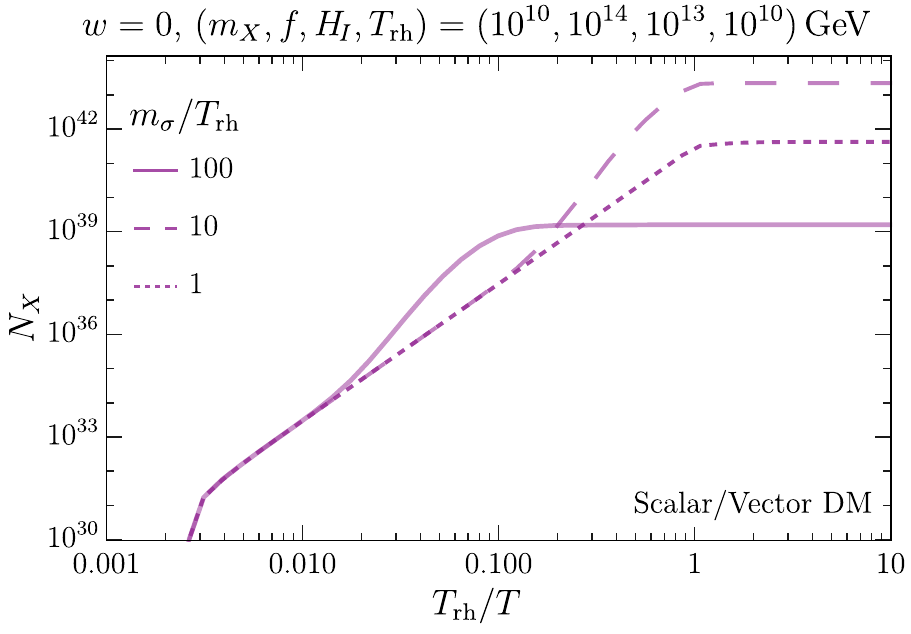}\!\!\!
\includegraphics[width=0.43\textwidth]{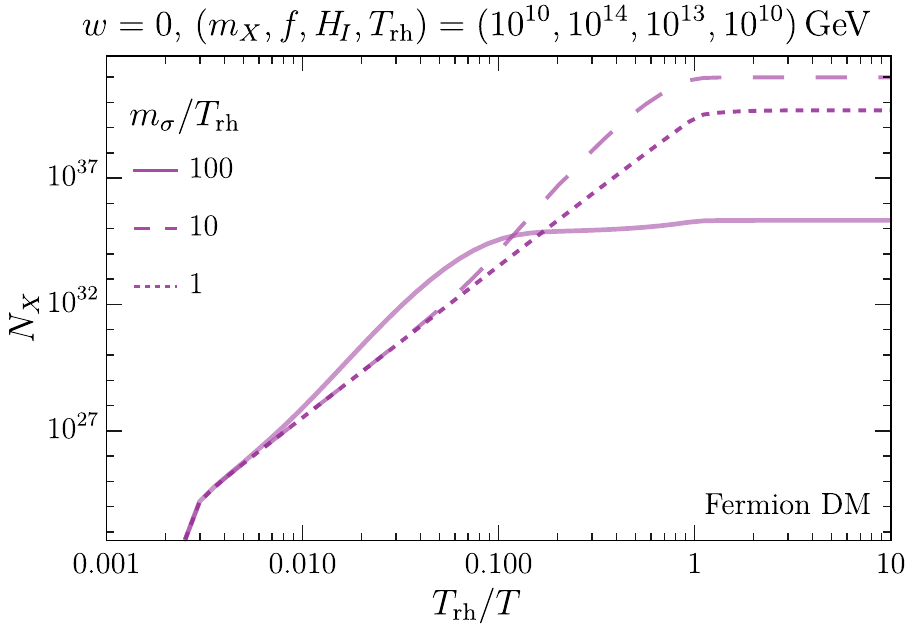}
\caption{Comoving number density $N_{\!X}$ for scalar/vector DM (left panel) and Dirac fermion DM (right panel) as a function of $T_{\rm rh}/T$ for different choices dilaton mass $m_\sigma$. }
\label{fig:NT_msdil}
\end{figure*}
\begin{figure*} [t!]
\centering
\includegraphics[width=0.43\textwidth]{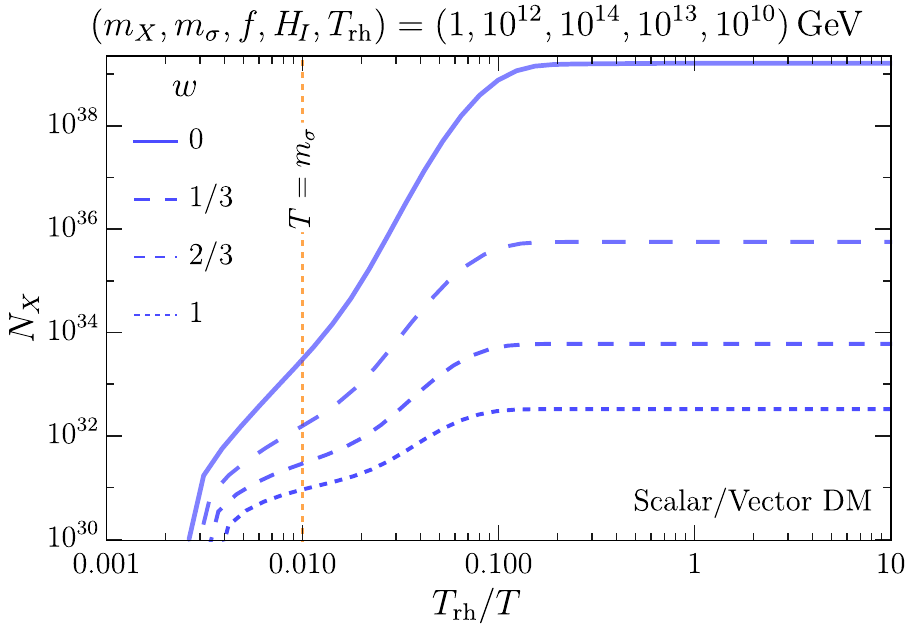}\!\!\!
\includegraphics[width=0.43\textwidth]{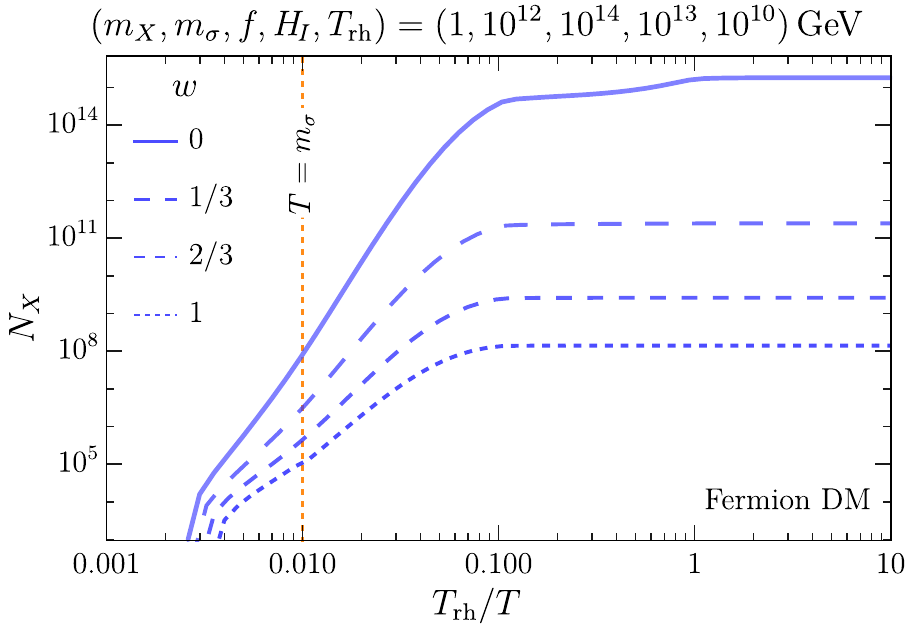}
\caption{Comoving number density $N_{\!X}$ for scalar/vector DM (left panel) and Dirac fermion DM (right panel) as a function of $T_{\rm rh}/T$ for different choices of the equation of state $w$. }
\label{fig:NT_w}
\end{figure*}

\subsubsection{Dependence on dilaton mass} 
For the case of scalar/vector DM (left panel) and fermion DM (right panel) in \fig{fig:NT_msdil}, we illustrate the dependence of comoving number density as a function of temperature for various values of dilaton mass $m_\sigma$ with fixed $m_{\!X}=T_{\rm rh}=10^{10}\gev$, $f=10^{14}\gev$, $H_{\! I}=10^{13}\gev$, and $w=0$.  We consider three choices for the dilaton mass $m_\sigma/T_{\rm rh}=100, 10, 1$, whereas for $m_\sigma/T_{\rm rh}<1$ the comoving number density remains almost the same as for the case $m_\sigma/T_{\rm rh}=1$. Again one can notice the scaling of $N_{\!X}$ with respect to $T$ for $T>m_\sigma$ where the dilaton can be produced on-shell and for the case when $T<m_\sigma$ where the dilaton can be effectively integrated out. 
Furthermore, due to naturalness one expects that the dilaton mass is of the same order as the conformal breaking scale $f$.

\subsubsection{Dependence on the equation of state during reheating} 
The nontrivial dependence on the equation of state $w$ during reheating is given in \eq{e.NX} as $\big(T_{\rm rh}/T\big)^{(7-w)/(1+w)}$. This shows suppression in the number density with increasing $w$.
In \fig{fig:NT_w} illustrates this feature where comoving number density $N_{\!X}$ is shown as a function of $T_{\rm rh}/T$ for $w=0,1/3,2/3,1$ with fixed values of $m_{\!X}=1\gev, m_\sigma=10^{12}\gev,f=10^{14}\gev$, and $T_{\rm rh}=10^{10}\gev$ for the scalar/vector DM (left panel) and fermion DM (right panel). Nonstandard cosmological effects are only relevant for temperatures $T<T_{\rm rh}$, i.e. during the reheating phase. 

\subsubsection{Dependence on the scale of inflation} 
\begin{figure*} [t!]
\centering
\includegraphics[width=0.43\textwidth]{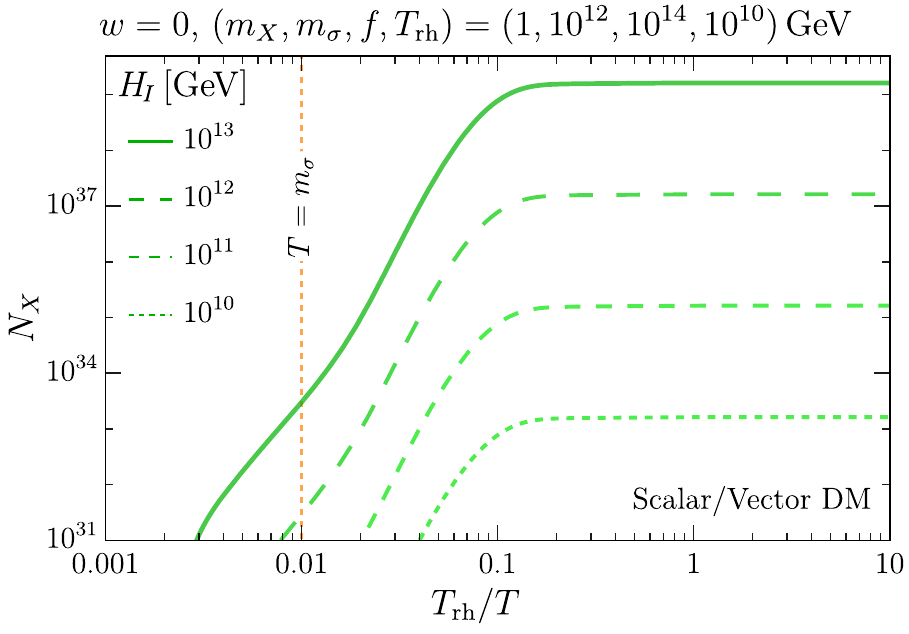}\!\!\!
\includegraphics[width=0.43\textwidth]{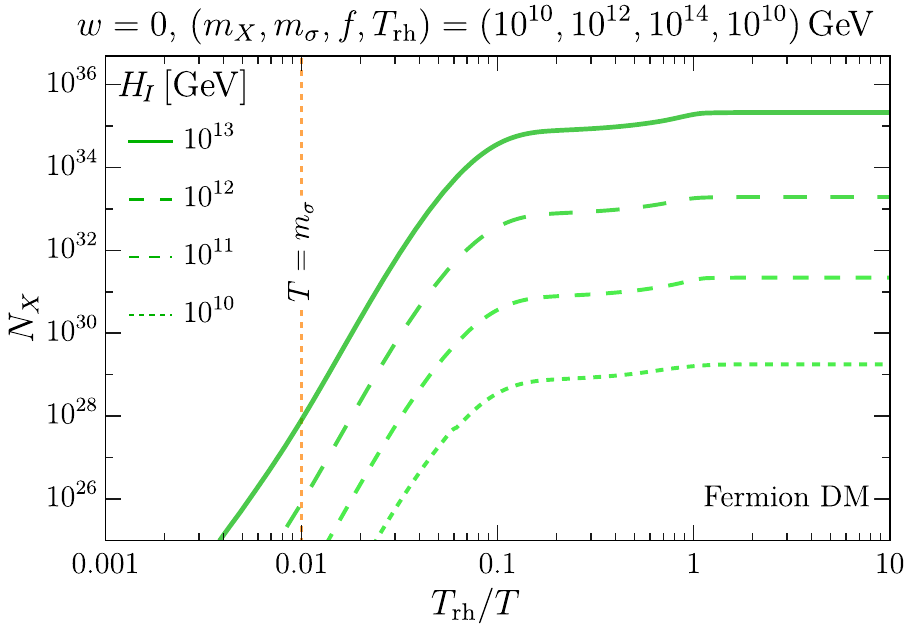}
\caption{Comoving number density $N_{\!X}$ for scalar/vector DM (left panel) and Dirac fermion DM (right panel) as a function of $T_{\rm rh}/T$ for different choices of the equation of state $w$. }
\label{fig:NT_Hi}
\end{figure*}

Dark matter production is sensitive to the scale of inflation $H_{\!I}$ which overall sets the scale of reheating dynamics. The dependence of $H_{\!I}$ on the comoving DM number density $N_X$ in Eq.~\eqref{e.NX} appears through $T_{\rm rh}$ and the maximum temperature during reheating phase $T_{\rm max}$. In \fig{fig:NT_Hi} we present the comoving DM number density $N_X$ as a function of $T_{\rm rh}/T$ for $H_{\!I}=(10^{13},10^{12},10^{11},10^{10})\gev$ for fixed values of $w\!=\!0, m_{\!X}\!=\!10^{10}\gev, m_\sigma\!=\!10^{12}\gev,f\!=\!10^{14}\gev$, and $T_{\rm rh}\!=\!10^{10}\gev$. The left panel shows the scalar/vector DM case, whereas the right panel represents the fermion DM case. Note that during the reheating phase, i.e. $T>T_{\rm rh}$, the $H_{\!I}$ dependence on DM production is nontrivial, however after the end of the reheating phase, i.e. $T<T_{\rm rh}$, the comoving DM number density is proportional to $H_{\!I}^2$. 

\subsubsection{Dark matter relic abundance} There are six free parameters in our model namely, $m_{\!X}$, $m_\sigma$, $f$, $H_{\! I}$, $T_{\rm rh}$, and $w$. In the following, we present regions of parameter space in the DM mass $m_{\!X}$ vs the reheating temperature $T_{\rm rh}$ with fixed choices of all the remaining parameters such that we produced to observed DM relic abundance $\Omega_{\!X} h^2=0.12$. In particular, in all the analyses we choose the Hubble scale at the end of inflation $H_e=10^{13}\gev$ and fix the ratio of dilaton mass to conformal breaking scale $m_\sigma/f=1\%$. Whereas the value of the conformal breaking scale fixes the observed DM relic abundance for a specific choice of the equation of state parameter $w$, which we take $w=0,1/3,2/3,1$.
\begin{figure*} [t!]
\centering
\includegraphics[width=0.33\textwidth]{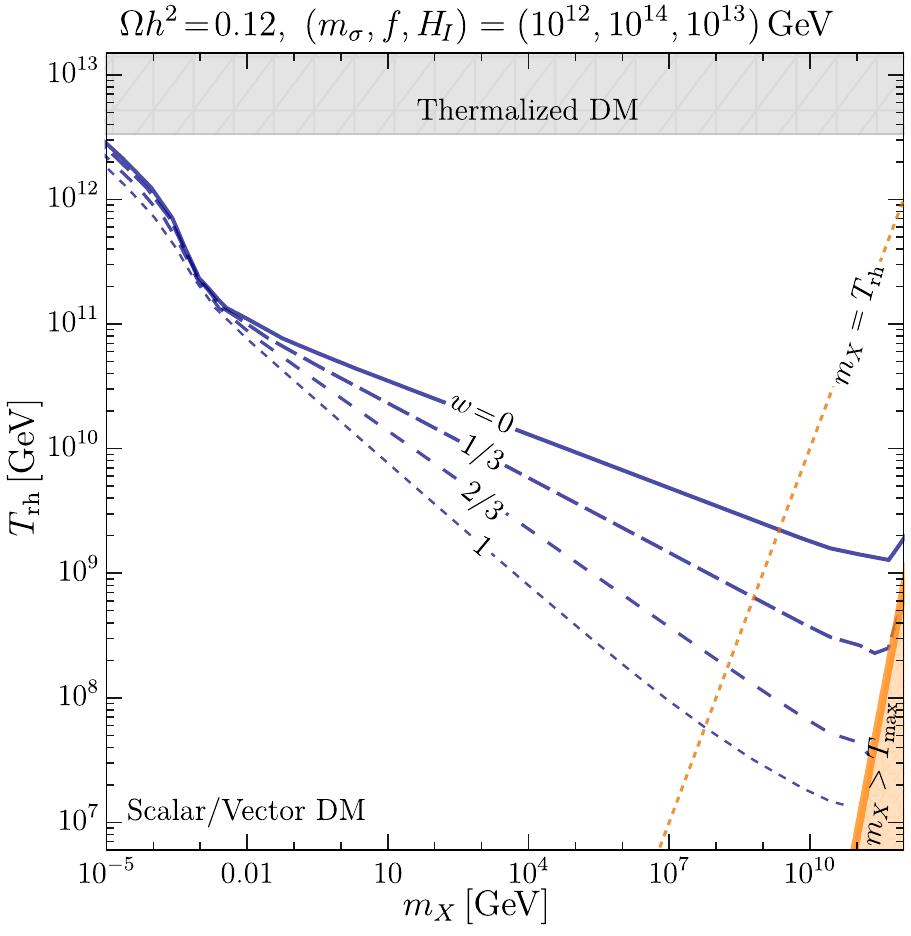}\!
\includegraphics[width=0.326\textwidth]{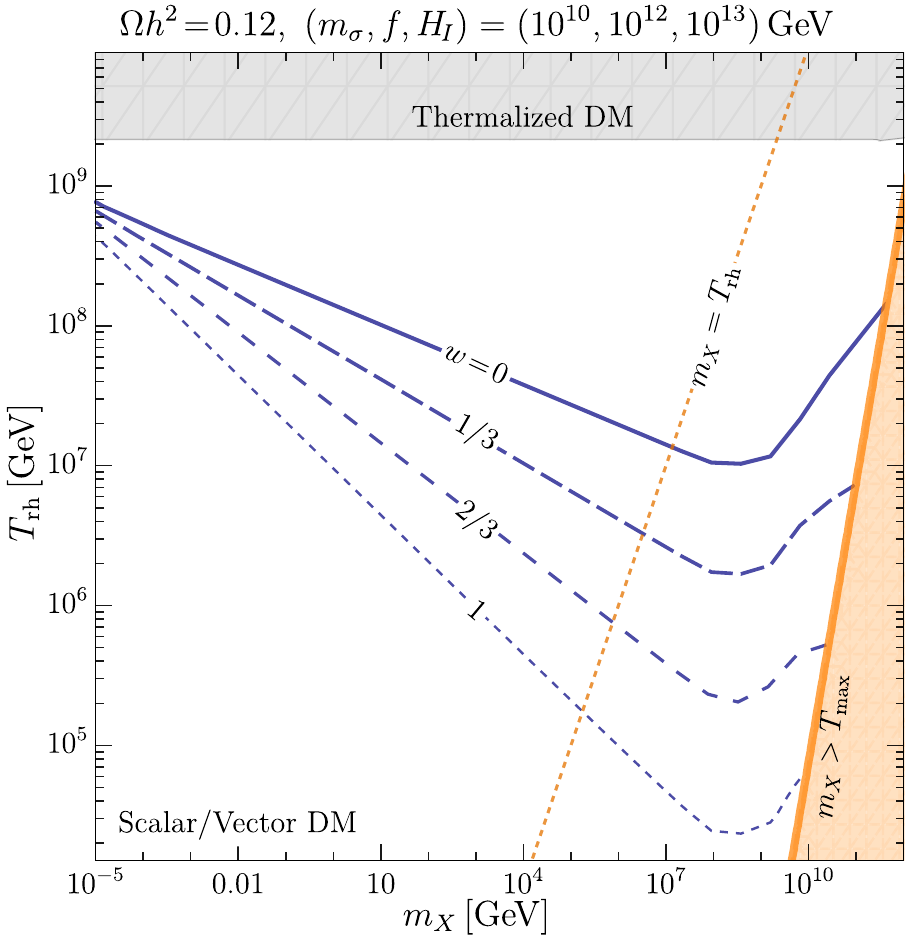}\!
\includegraphics[width=0.33\textwidth]{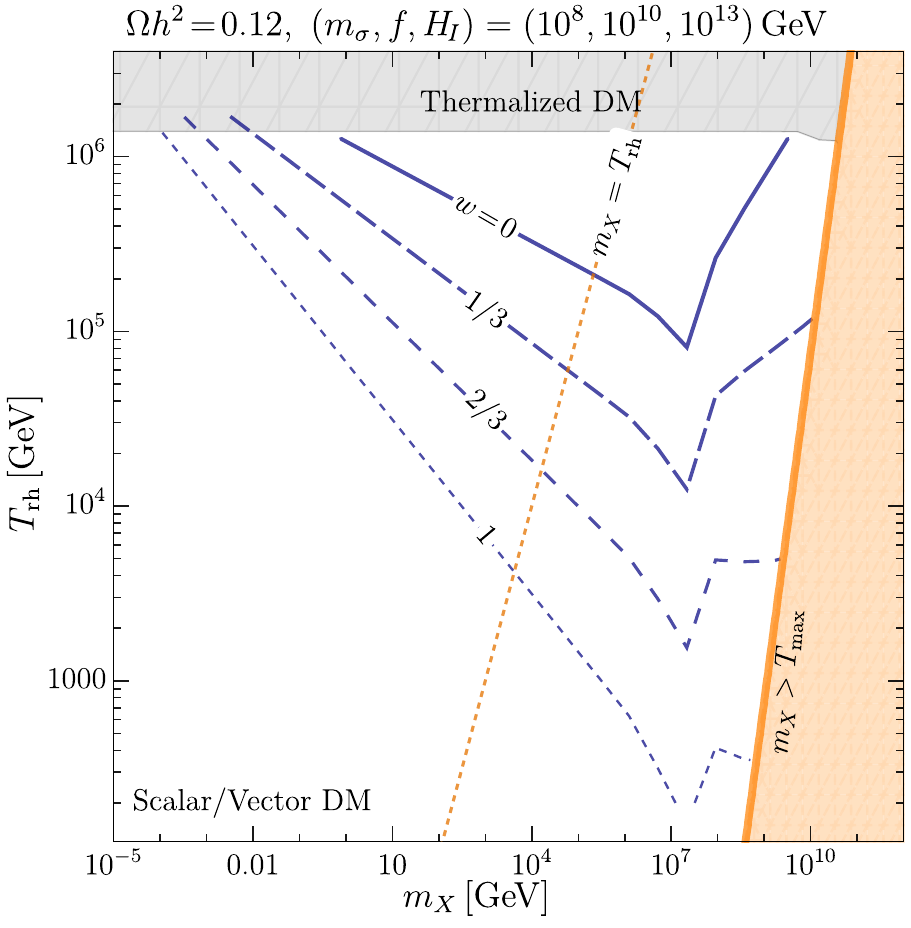}
\caption{Contours of observed DM relic abundance for scalar/vector DM as a function of $m_{\!X}$ and $T_{\rm rh}$ for different choices of the equation of state parameter $w$. }
\label{fig:mxTrh_svec}
\end{figure*}
\begin{figure*} [t!]
\centering
\includegraphics[width=0.33\textwidth]{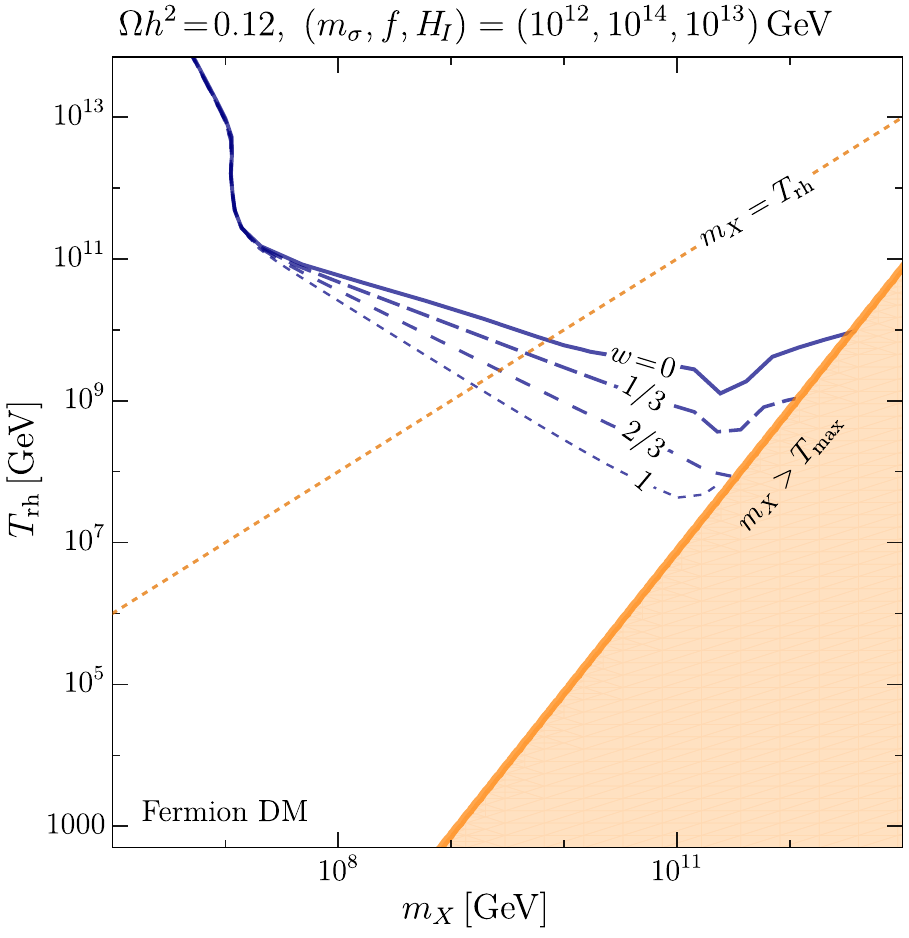}\!
\includegraphics[width=0.33\textwidth]{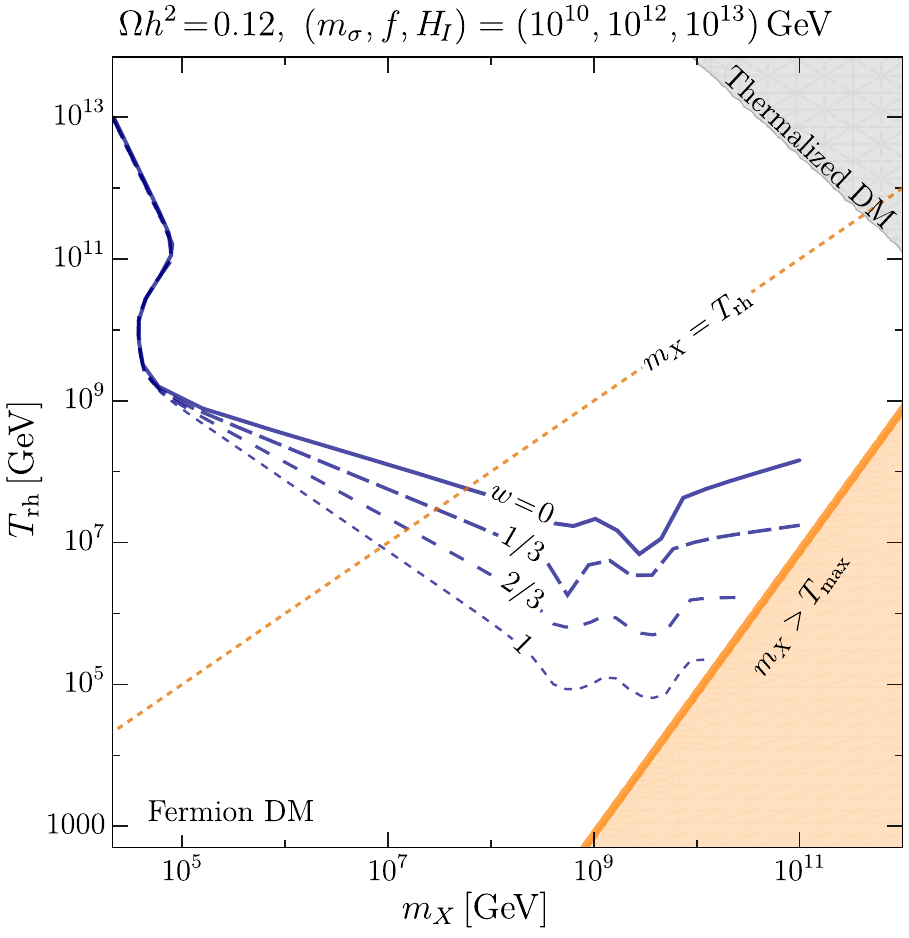}\!
\includegraphics[width=0.33\textwidth]{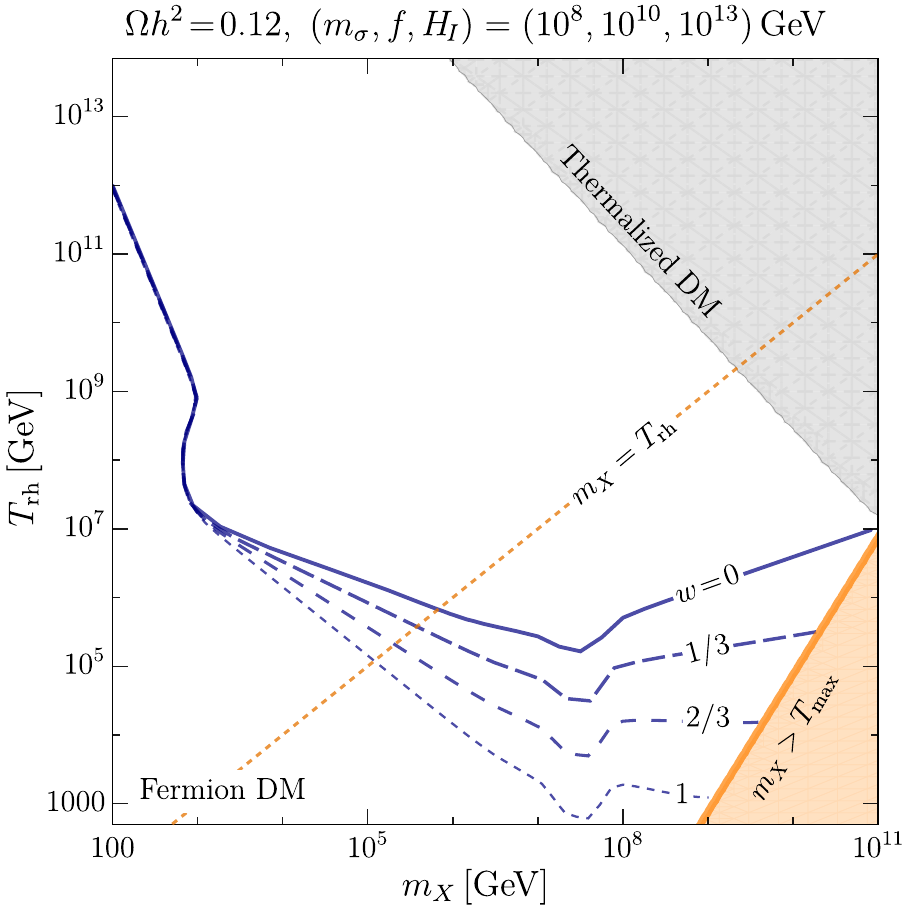}
\caption{Contours of observed DM relic abundance for fermionic DM as a function of $m_{\!X}$ and $T_{\rm rh}$ for different choices of the equation of state parameter $w$. }
\label{fig:mxTrh_ferm}
\end{figure*}

In \fig{fig:mxTrh_svec}, we present contours of observed DM relic abundance for the case of scalar/vector DM in the $T_{\rm rh}$ vs $m_\chi$ plane for different values of the equation of state $w$ and $f$ with fixed values of $m_\sigma/f=1\%$ and $H_{\! I}=10^{13}\gev$. The gray shaded region represents parameter space where the DM would be in thermal equilibrium with the SM bath. The orange shaded region represents the parameter space where $m_{\!X}>T_{\rm max}$ and dashed-orange line denotes $m_{\!X}=T_{\rm rh}$. The three panels of this figure show the results for conformal breaking scale $f=10^{14}, 10^{12}, 10^{10}\gev$ from left to right, respectively. The curves show observed DM abundance for DM masses from $10^{-6}\gev$ to $10^{12}\gev$ as a function of the reheating temperature. Note that for a low DM mass $m_\chi$ we require larger $T_{\rm rh}$ to produce desired DM abundance, whereas for a high DM mass $m_\chi$ we require smaller $T_{\rm rh}$. The lower values of $T_{\rm rh}$ imply that the reheating phase lasts longer and therefore, the impact of nonstandard cosmology during this phase has a significant effect on the DM production. This effect is manifestly shown in \fig{fig:mxTrh_svec}. 
Note that the dip in these curves for large DM masses is around $m_{\!X}\sim m_\sigma$ as the temperature dependence of the source term changes as shown in \eq{eq_CX_scalar} for the cases of scalar/vector DM.

In \fig{fig:mxTrh_ferm} we show the fermionic DM results for parameters in $m_{\!X}-T_{\rm rh}$ plane which produces observed DM relic abundance for different choices of $w$ and $f$, whereas we fix $m_\sigma/f=1\%$ and $H_{\! I}=10^{13}\gev$. Similar to \fig{fig:mxTrh_svec}, in \fig{fig:mxTrh_ferm} the gray shaded region represents the DM in thermal equilibrium with the SM bath and the orange shaded region shows the DM mass larger than the maximum temperature of the thermal bath, i.e. $m_{\!X}>T_{\rm max}$. Note that in the case of fermionic DM, unlike the bosonic DM case, the scattering cross section is proportional to DM mass squared. Therefore the DM thermalization condition is proportional to DM mass, for larger masses, the cross section is large and hence it is thermalized, whereas the constraint weakens for lower DM masses. Furthermore, due to the same DM mass dependence notice that for larger values of conformal breaking scaling, e.g. $f=10^{14}\gev$ (left panel) the observed DM abundance can only be achieved for DM mass $m_{\!X}>10^6\gev$, in contrast to the scalar/vector DM case where sub-GeV masses are also allowed. Note that the effect of nonstandard cosmology during the reheating phase is very similar to that of the scalar/vector DM case, i.e. for smaller values of reheating temperature the duration of the reheating phase is larger, and therefore the effect of the equation of state $w$ during this phase is significant. However, for larger reheating temperatures the effect is irrelevant. 

\section{Conclusions}
\label{sec:conc}
In this work, we have studied the implication of nonstandard cosmology during reheating on the ultraviolet freeze-in production of DM via the dilaton portal. 
We assume the SM and DM are part of a (strongly coupled) conformal/scale-invariant theory, where the scale invariance is broken spontaneously at scale $\Lambda\!=\!4\pi f$. As a result, the low energy effective theory contains a dilaton field $\sigma$ which couples to the SM and DM through higher dimensional operators suppressed by the breaking scale. 
Furthermore, we assume DM interacts with the SM only through the dilaton portal. 
In this framework, the lowest dimensional interaction between the SM and DM is a dimension-six operator. 
For large values of conformal breaking scale $f$ such interactions are naturally very small and as a result, the DM is not in thermal equilibrium with the SM bath.  
Therefore, DM can only be produced in the early universe through the freeze-in mechanism. 
There are two production processes through which the DM can be produced: (i) through the annihilation of SM particles to DM, and (ii) through direct decays of dilaton field to DM when the dilaton mass is larger than twice the DM mass. We studied the cases when the DM is a scalar, vector, or fermion field. 

The ultraviolet freeze-in production of DM is highly sensitive to the maximum temperature of the SM bath particles as well as the reheating dynamics. 
We have parametrized the reheating dynamics with three parameters, (a) Hubble scale at the end of inflation $H_{\! I}$ which determines inflaton energy density at the end of inflation, (b) the equation of state $w$ of the inflaton field during reheating, and (c) the reheating temperature $T_{\rm rh}$ when the inflaton energy density is equal to that of the SM, therefore it determines the duration of reheating. The dilaton portal dynamics involve two parameters, the dilaton mass $m_\sigma$, and the conformal breaking scale $f$. Apart from these five parameters, DM mass $m_{\!X}$ (for scalar/vector/fermion) is the only remaining free parameter. 
In this study, we considered a high-scale inflationary scenario, where the Hubble scale at the end of inflation is fixed to be $H_{\! I}=10^{13}\gev$. With this choice of $H_{\! I}$, assuming instantaneous thermalization the maximum temperature attained by the SM bath is $T_{\rm max}\sim 10^{15} \gev\times\sqrt{T_{\rm rh}/(10^{15} \gev)}$, see \eq{eq:Tmax}. For the equation of state parameter, we choose $w=0,1/3,2/3,1$, where $w=0$ and $w=1/3$ correspond to the matter-dominated and radiation-dominated reheating which are achieved for quadratic and quartic inflaton potentials, respectively. 
We have shown the dependence of the DM number density as a function of temperature on various parameters of the model in Figs.~\ref{fig:NT_mX}--\ref{fig:NT_Hi}. 

In \fig{fig:mxTrh_svec} and \fig{fig:mxTrh_ferm}, we present the results in the $T_{\rm rh}-m_{\!X}$ plane for the scalar/vector and fermion DM cases, respectively. We consider high-scale conformal breaking scale $f=10^{14},10^{12},10^{10}\gev$ and we fix the dilaton mass $m_\sigma/f=1\%$. 
It is shown that DM production is sensitive to the conformal breaking scale $f$ as the production cross section scales as $1/f^4$. To realize DM freeze-in production mechanism, one requires high-scale conformal breaking. For the case of scalar/vector DM (\fig{fig:mxTrh_svec}), the observed DM abundance can be achieved for DM masses in the range $10^{-5}-10^{12}\gev$ depending on the reheating temperature $T_{\rm rh}$ and breaking scale $f$. Whereas, for the case of fermion DM (\fig{fig:mxTrh_ferm}), the observed DM abundance can be achieved for DM masses in the range $10^{2}-10^{12}\gev$ depending on the reheating temperature $T_{\rm rh}$ and breaking scale $f$. We conclude that the dilaton portal offers a natural realization of ultraviolet freeze-in production of DM for a wide DM mass range and its sensitivity to the reheating dynamics is investigated. 

\section*{Acknowledgements}
The research of SN is supported by the Cluster of Excellence {\it Precision Physics, Fundamental Interactions and Structure of Matter} (PRISMA$^+$ -- EXC 2118/1) within the German Excellence Strategy (project ID 39083149).


\bibliography{bib_FIDP}{}
\bibliographystyle{aabib}

\end{document}